\documentclass{pseudo-ws-ijqi}

\usepackage{amssymb}
\usepackage{setspace}
\usepackage [sort,superscript,biblabel] {cite} 
\usepackage{multirow}
\usepackage{colordvi}

\newcommand{\ket}     [1] {\ensuremath{\left|#1\right\rangle}}

\newcommand{\iso} [2] {\ensuremath{{}^{#2}\!{\mathrm{#1}}}}
\newcommand{\multiso} [3] {\ensuremath{{}^{#2}\!{\mathrm{#1}_{#3}}}}

\begin{document}

\markboth{Yosi Atia, Yuval Elias, Tal Mor, and Yossi Weinstein} 
{Quantum Computing Gates via Optimal Control}

\catchline{}{}{}{}{}

\title{Quantum Computing Gates via Optimal Control}

\author{ATIA, YOSI\footnote{Corresponding author.}}
\address{Computer Science and Engineering, The Hebrew University of Jerusalem,  Israel 91904, yosiat@cs.huji.ac.il}

\author{ELIAS, YUVAL}
\address{Computer Science Dept., Technion - Israel Institute of Technology, 
Technion City, Haifa, Israel 3200008, ye1@cs.technion.ac.il}

\author{MOR, TAL}
\address{Computer Science Dept., Technion - Israel Institute of Technology, 
Technion City, Haifa, Israel 3200008, talmo@cs.technion.ac.il}

\author{WEINSTEIN, YOSSI}
\address{Computer Science Dept., Technion - Israel Institute of Technology, 
Technion City, Haifa, Israel 3200008, yossiv@cs.technion.ac.il}

\maketitle

\begin{history}
\received{30 Aug 2013}
\end{history}


\begin{abstract}
We demonstrate the use of optimal control to design two entropy-manipulating
quantum gates which are more complex than the corresponding, commonly used, gates, such as CNOT and Toffoli (CCNOT):
A 2-qubit gate called PE (polarization exchange) and a 3-qubit gate called COMP (polarization compression) were designed using GRAPE, an optimal control algorithm. Both gates were designed for a three-spin system. Our design provided efficient and robust NMR radio frequency (RF) pulses for \multiso{C}{13}{2}-trichloroethylene (TCE), our chosen three-spin system.
We then experimentally applied these two quantum gates onto TCE at the NMR lab.
Such design of these gates and others could be relevant for near-future applications of quantum computing devices.
\end{abstract}


\section{Introduction}
One of the major challenges in building a quantum computer is to coherently control a large quantum system well enough to perform an arbitrary quantum algorithm.\cite{DiVincenzo00} Nuclear magnetic resonance (NMR) offers an excellent test bed for developing techniques to control quantum systems.\cite{RLL09} Applying a quantum algorithm in NMR requires RF pulses that manipulate the spin system through the various steps of the algorithm.

In NMR, optimal control is used for designing and optimizing experiments for various applications such as imaging,\cite{CSNDMA+86,MMSA+69,RZ+96,XKZML+08} liquid and solid state spectroscopy,\cite{Khaneja2005,RKG+02,KRLG+03,KSBRKSN+04,KVKGN+05,VKKSN+05,SIMPSON+08} quantum computation,\cite{ZYK+08,KGB+02,SSKG+05,KHSYSG+07} dynamic nuclear polarization (DNP),\cite{MTZN+08,PHKG+08,HYRC+08,Khaneja+07} and algorithmic cooling in the solid state.\cite{BLMR+08} 
Pulses in NMR can be designed by SIMPSON (see Ref.~\refcite{SIMPSON+00}), which is, originally, an NMR simulation program. SIMPSON was lately expanded to implement the optimal control algorithm 
GRAPE.\cite{Khaneja2005,SIMPSON+08,RNLKL08,NKGK10,RLL09,RJ12,CPB12,FLL12}

In this work, we designed, using SIMPSON, 
a polarization exchange (PE) pulse, on a 3-qubit system.  
We then also implemented that designed pulse experimentally.
A second pulse, compression (COMP), which requires more time hence
is more challenging, was also designed and implemented on the same experimental system. 
These gates were chosen since they manipulate (redistribute) 
the entropy of two or more spins; PE is important in NMR in general,
and COMP performs 3-bit reversible entropy compression.\cite{Sorensen89,SV99}

These two gates and similar ones are building blocks of 
algorithmic cooling (see Refs.~\refcite{BMRVV02,FLMR04}), hence similar gates were already designed for that purpose. 
The first attempts were based on directly designing a unitary operation, with no numerical optimization. 
See for instance Ref.~\refcite{CVS01}, for compression, and Ref.~\refcite{POTENT} 
for polarization transfer (and heat-bath cooling beyond Shannon's entropy bound).
However,
these initial attempts suffered from a large error in the experimentally implemented gates.  
The first full algorithmic cooling experiment
in solid state NMR, described in Ref.~\refcite{BMR+05}, utilized numerical optimization to design COMP and SWAP pulses for \iso{C}{13}-labeled malonic acid. Later, GRAPE was applied to this system
to obtain higher pulse fidelity, permitting four rounds of algorithmic cooling (beyond the polarization of the heat-bath).\cite{BLMR+08} Motivated by this result, the
optimized pulses designed here have recently been combined to demonstrate
algorithmic cooling in liquid state NMR\cite{AEMW14,AtiaMSc} (based on the work presented here). 
GRAPE has also been used to design pulses for up to seven qubits in crotonic acid.\cite{RNLKL08,FLL12}

The rest of this work consists of the following sections:
Chapter~\ref{chap:OC4QGate} highlights the advantages of the state-to-state optimal
 control approach for the design of quantum gates.
Chapter~\ref{chap:materials-and-methods} describes the experimental apparatus and the process of pulse design and analysis. 
The designed optimized pulses and the resulting spectra are described in Chapter~\ref{chap:Results}. 
We discuss the results and future prospects in Chapter~\ref{chap:Discussion}.

\section{Optimal Control as a Tool for Designing Quantum Gates}\label{chap:OC4QGate}
In this section we briefly clarify the advantages of state-to-state design of
 quantum gates by means of the well-established optimal control methodology. We
 focus on the gates relevant for this paper, polarization exchange and compression.

\subsection{Numerical Optimization of the Quantum Gates}\label{sec:Numerical-Opt-QGates}
As mentioned above, researchers originally 
tried to implement a chosen unitary gate directly, with no numerical optimization.\cite{CVS01,POTENT} 
However, due to the need to compansate for undesired free evolution under Ising 
or Heisenberg spin-coupling and chemical shifts, the resulting pulse sequences were quite long and each pulse added a non-negligible noise.
For example, the compression is supposed to enhance the polarization from
100\% to 150\%, and in Ref.~\refcite{CVS01} the obtained result was only around
122\%. Similarly, multiple PE gates were applied in Ref.~\refcite{POTENT}, and one of them had a fidelity of around 70\% only. Numerical optimization allowed better results (see Ref.~\refcite{BMR+05}) but their design was rather complicated and a better method was required.\footnote{Related results were obtained for two-qubit gates involving superconducting qubits, using a numerical optimization approach based on simulated annealing.\cite{LCP08}}  

State-to-state optimal control method has the advantage of fully using the redundancy allowed in gate design. Instead of choosing a unitary transformation, the designer only tells the optimization software (SIMPSON) the initial state and the desired final state, while giving the software full freedom in searching for the optimal pulse. This freedom is then somewhat limited by fixing
the duration of the pulse, and by demanding robustness of the pulse to small changes.  

\subsection{Designing polarization exchange and compression using the state to state approach}\label{sec:PE-COMP-Gates}

A SWAP can be implemented via three controlled-NOT gates, in case these are easier to implement. 
In NMR, applying a specific gate to two spins while doing nothing to all other
spins is highly challenging (because spin coupling and chemical shifts are continuously present). 

Polarization Exchange (PE) ``gate'' is a variant of SWAP in which we only want to interchange the z-components of the spins without caring about the phases, thus we have some phase redundancy. For example, equation~\ref{eq:SWAP-variant} displays a variant of SWAP which does not perform a good swap of the states, but is perfect for PE:  
\begin{eqnarray} \label{eq:SWAP-variant}
\ket{00} \rightarrow  \phantom{-}\ket{00} \nonumber  \\
\ket{01} \rightarrow  \phantom{-}\ket{10} \nonumber  \\
\ket{10} \rightarrow  \phantom{-}\ket{01} \nonumber  \\
\ket{11} \rightarrow   -\ket{11}{}
\end{eqnarray}
(note the phase in the last line). Thus, PE is actually not a single gate but a
family of gates. 

Another source of redundancy arises when two z-states have the same probability, for example if the density matrix is diagonal, such as $diag(3,1,1,-1)/4$ then any unitary transformation inside the subspace spanned by $\ket{01}$ and $\ket{10}$ does not disturb the PE, hence this gives another source of redundancy. This opportunity can appear anywhere during the pulse.

Any unitary operator $U$ (obtained by applying a fixed Hamiltonian for some time) can be replaced by a different trajectory, e.g., $U=U_1U_2$ even if $U_1$ is arbitrary. This is a source of redundancy which is independent of the above sources; for example, both SWAP and the variant of SWAP from eq.~\ref{eq:SWAP-variant} can be applied by an infinite variety of implementations.

Due to all these sources of redundancy the variety of potential pulses is expected to be quite vast, especially if we use the maximal incrementation allowed by the software. Thus there is a very good probability of finding robust pulses.

We want our pulse to be short, in order not to accumulate dissipation or errors. This limits the variety of redundant pulses the optimization can choose from; the redundancy increases with the number of increments.

Compression on three spins puts the maximal possible polarization on one of them. 
Similar to PE, compression is a family of gates, because there is a redundancy in terms of the remaining polarization on the other spins. Even if we limit that
redundancy, by maximizing first the polarization on one spin, and then on the
next (while keeping the maximum on the first one of course), etc., still there
is redundancy in terms of the phase, trajectory and possibly identical probabilities of z-states.
Here is a compression gate that we call COMP:
$\ket{011} \leftrightarrow \ket{100}.$
See Ref.~\refcite{FLMR04} for explanation of the compression performed using this gate and similar ones.

\section{Materials and Methods}\label{chap:materials-and-methods}

\begin{figure} [!h]
		\centering
			\includegraphics [scale=0.5] {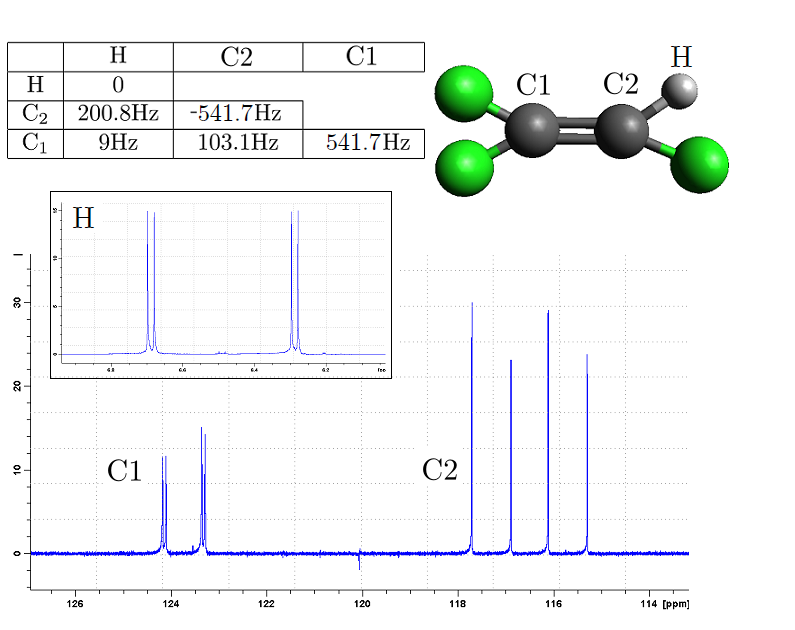}

\caption [TCE, its spin system properties and spectra.] { Three spins in \multiso{C}{13}{2}-TCE were utilized in the experiment: H, C2, and C1. In the table, the chemical shifts relative to the transmitter frequency are in the diagonal elements, and the J-couplings are in the off diagonal elements. Note that the broadcasting frequency of the carbon channel is between the Larmor frequencies of C1 and C2. The carbon spectrum is at the bottom, the proton spectrum is in the small frame.   
Remark: The carbons have different T$_2^*$, hence their peaks have different heights, although their integrals, and therefore their polarizations, are equal up to an error of 2\%.}						
		\label{fig:TCE_with_spectrum_ch3}
	\end{figure}

The experiments were performed on a Bruker Avance III 600 spectrometer using a standard 5~mm BBO probe. The sample was \multiso{C}{13}{2}-TCE  with paramagnetic reagent Cr(acac)$_3$ for increased $T_1$ ratios (see Ref.~\refcite{FMW05}) in CDCl$_3$ (chloroform-d) solution, comprising two  \iso{C}{13} and one \iso{H}{1} qubits (see Figure \ref{fig:TCE_with_spectrum_ch3}). 

\subsection{GRAPE - GRadient Ascent Pulse Engineering}
GRAPE\cite{Khaneja2005} is an optimal control algorithm which designs shaped pulses contemplating to apply a state to state transformation, or a given unitary propagator. A pulse with random shape (i.e., random amplitudes and phases) is generated, and then monotonically improved by gradient ascent\footnote{Not to be confused with a magnetic field gradient used in NMR experiments.} in respect to a target function. Robustness of the pulses to experimental parameters such as RF or primary magnetic field inhomogeneities is supported inherently by the algorithm. In this work we used the GRAPE implementation in SIMPSON.\cite{SIMPSON+08} 

The GRAPE state-to-state algorithm seeks the optimal amplitudes and phases of the RF fields that transform a given initial density matrix \(\rho(0)=\rho_{0}\) sufficiently close to a desired target density matrix \(\rho_{desired}\). For a specified pulse duration $T$, GRAPE attempts to increase the overlap between \(\rho(T)\) and \(\rho_{desired}\) beyond a given threshold.

The total energy of the pulse might be limited by hardware or the sample allowed temperatures. GRAPE can consider physical constraints such as the the maximum intensity of the RF transceiver, and can enable robustness to deviation e.g., in the RF pulse homogeneity. See table~\ref{SensTable}.

\subsection{Measurement of system parameters.}

SIMPSON requires the full Hamiltonian of the system, namely, the J-coupling constants and the chemical shifts of the spins. J-coupling constants were found directly from the spins' excitation  (90$^\circ$) spectra, i.e., by using Bruker \textbf{zg} pulse program. 
However, for the carbons the Larmor frequency difference is not much greater than the J-coupling, therefore it does not provide a good approximation for the chemical shift.
We preferred to calculate their chemical shifts by matching the measured spectrum in the lab to the predicted spectrum of TCE using SIMPSON in simulation mode. After collecting the parameters measured, the rotating frame Hamiltonian used for the optimization is (coupling constants and chemical shifts are in Hz):

\begin{equation}
H=2\pi\hbar\left(541.7 I_z^{C1}-541.7 I_z^{C2}+200.8I_z^{C2}I_z^H+103.1I_z^{C1}I_z^{C2}+9I_z^{C1}I_z^{H}\right)
\end{equation}

\renewcommand{\arraystretch}{1.5}  
\begin{table}[h!]
\tbl{Estimated sensitivity of the compression pulse to parameters.
The expected fidelity for nominal parameters is 0.987.} {
\begin{tabular}{|p{5cm}|c|c|c|}
\hline 
Parameter  & Nominal Value  & Deviated Value  & Expected Fidelity\tabularnewline
\hline 
\multirow{2}{5cm}{Hydrogen resonance frequency (Principle magnetic field inhomogeneity) } & \multirow{2}{*}{600.55~MHz } & 606.13~MHz  & 0.9958\tabularnewline
\cline{3-4} 
 &  & 594.55~MHz  & 0.9956\tabularnewline
\hline 
\multirow{2}{5cm}{$J_{H,C_{2}}$ } & \multirow{2}{*}{201~Hz } & 221~Hz  & 0.9862\tabularnewline
\cline{3-4} 
 &  & 180~Hz  & 0.9841\tabularnewline
\hline 
\multirow{2}{5cm}{$J_{C_{1},C_{2}}$ } & \multirow{2}{*}{103~Hz } & 113~Hz  & 0.9912\tabularnewline
\cline{3-4} 
 &  & 93~Hz  & 0.9891\tabularnewline
\hline 
\multirow{2}{5cm}{$J_{H,C1}$ } & \multirow{2}{*}{9~Hz } & 12~Hz  & 0.9983\tabularnewline
\cline{3-4} 
 &  & 5~Hz  & 0.9984\tabularnewline
\hline 
\multirow{2}{5cm}{Pulse duration } & \multirow{2}{*}{13~msec } & 13.2~msec  & 0.9918\tabularnewline
\cline{3-4} 
 &  & 12.8~msec  & 0.9907\tabularnewline
\hline 
\multirow{2}{5cm}{Hydrogen frequency miscalibration} & \multirow{2}{*}{0 } & +20~Hz  & 0.9887\tabularnewline
\cline{3-4} 
 &  & -20~Hz  & 0.9911\tabularnewline
\hline 
\multirow{2}{5cm}{Carbon frequency miscalibration } & \multirow{2}{*}{0 } & +20~Hz  & 0.9735\tabularnewline
\cline{3-4} 
 &  & -20~Hz  & 0.9692\tabularnewline
\hline 
\multirow{2}{5cm}{Difference between the carbons' Larmor frequencies} & \multirow{2}{*}{1083~Hz } & 1103~Hz  & 0.9922\tabularnewline
\cline{3-4} 
 &  & 1063~Hz  & 0.9917\tabularnewline
\hline 

\end{tabular}}
\label{SensTable}
\end{table}
\renewcommand{\arraystretch}{1}  

Once a pulse is generated it is possible to see how sensitive it is to various parameters by evaluating the target function of the pulse with parameters deviating from the nominal.\footnote{Sensitivity that originates from thermalization or dephasing will not be accurately estimated since SIMPSON does not take them into account. Also, sensitivity to a combination of variables was not tested.}
Knowing the sensitivity of a parameter is utilized for determining how much effort should be made in tuning or measuring it. The required robustness of the pulse to this parameter is also deduced from the sensitivity. Table \ref{SensTable} summarizes the sensitivity to different parameters of the compression pulse, with expected nominal fidelity of 0.9985.  

Table \ref{tbl:TCET1T2} summarizes the T$_1$ and T$_2^*$ values of the spin system.
T$_2^*$ was measured directly from the spectrum (single scan, line-width at half maximum). T$_1$ of each spin was measured in a different experiment, a few days earlier, by a standard inversion recovery method (Bruker pulse program: \textbf{t1ir}\footnote{In \textbf{t1ir} (inversion recovery) sequence, a 180$^\circ$ pulse is applied, followed by a varying delay $\tau$, a 90$^\circ$ pulse and a data acquisition.}). Typically, 17 logarithmically even spaced delays were used, and T$_1$ was found by fitting the data to the expected exponential model. 

\begin{table}[h]
\tbl{Measured T$_1$ and T$_2^*$ values of TCE.}
{\begin{tabular}{@{}cccc@{}} \toprule
& H & C2 & C1 \\ \colrule
T$_1$~(sec)\hphantom{00} & \hphantom{0}$2.67\pm0.03$ & \hphantom{0}$17.3\pm0.2$  & \hphantom{0}$29.2\pm0.1$ \\
T$_{2}^*$~(msec)\hphantom{00} & \hphantom{0}$200\pm10$ & \hphantom{0}$440\pm30$ & \hphantom{0}$230\pm10$ \\ \botrule
\end{tabular}}
\label{tbl:TCET1T2}
\end{table}

\subsection{Pulse design\label{sec:PulseDesignDetails}}

The PE pulse was generated by SIMPSON according to the following input parameters:
\begin{itemize}
	\item \textbf{Initial state:}  $I_z^{C1}+I_z^{C2}+4I_z^H$ $\propto$ diag$(6,-2,4,-4,4,-4,2,-6)$
	\item \textbf{Final state:} $I_z^{C1}+4I_z^{C2}+I_z^H$ $\propto$ diag$(6,4,-2,-4,4,2,-4,-6)$
	\item \textbf{Max RF:} 2~kHz
	\item \textbf{Duration:} 6~msec
\end{itemize}

 The compression pulse was generated with the following SIMPSON parameters:
{\raggedright
\begin{itemize}
	\item \textbf{Initial state:} $I_z^{C1}+I_z^{C2}+I_z^H$ $\propto$ diag$(3,1,1,-1,1,-1,-1,-3)$
	\item \textbf{Final state:} $\frac{3}{2}I_z^{C1}+\frac{1}{2}I_z^{C2}+\frac{1}{2} I_z^H + 2I_z^HI_z^{C2}I_z^{C1} $ $\propto diag( 3, 1, 1, 1, -1, -1, -1, -3 )$

	\item \textbf{Max RF:} 2~kHz
	\item \textbf{Duration:} 13~msec
\end{itemize}
}
Notice that in both experiments the final and initial states are degenerate allowing redundancy of potential pulses.

The computation of the pulses required hundreds of iterations of the optimization, and hours or even days of computer time, depending on the complexity of the pulse and its constraints. In order to reduce the calculation time, the pulse was first optimized without RF robustness, to reach the vicinity of the efficient pulses.  Then, a robustness constraint was added to the optimization, improving the pulse further (but vastly increasing the calculation time of each iteration).

Although infinite number of possible compression pulses maximize the polarization of C1, several considerations were made when selecting the final density matrix. It is desirable that at the end of the compression, C2 and H would retain some of their polarization, in order to reduce the reset times of the hydrogen in the next steps of algorithmic cooling. Non-required density matrix terms with long thermalization time  were avoided as they don't vanish before the next PE step, and may hinder the cooling. Note that the coupling constants determine the transition times, hence choosing a final density matrix sets a lower bound on the pulse length, indirectly affecting the efficiency. 
We chose a final density matrix which is identical to the result of applying 3-bit-compression (see Ref.~\refcite{FLMR04}) to the initial density matrix. 

SIMPSON does not take dissipation mechanisms into account. This is one of the reasons the observed efficiency of the designed pulses in the lab is lower than predicted by SIMPSON\@.
The dissipation rates of a system depend on its state. For example, when a spin is on the x axis it will experience both $T_1$ and $T_2$ relaxation, while on the z axis it will experience only $T_1$ relaxation. Pulses which apply the same unitary operator might also evolve the system in different paths that will cause a different amount of dissipation, and will therefore have different fidelities. Therefore, the uncertainty of the pulse fidelity in the lab was dealt with by generating several pulses that apply the same operation (state to state), and picking the pulse with best fidelity in the lab. In future work, an analysis of dissipation processes can be done in order to find an optimal path. A better pulse could then be prepared by tailoring multiple pulses (generated by SIMPSON) that evolve the system through states in the optimal path.

\subsection{Reference spectrum and power calibration}
Before each lab session, two key procedures were performed: acquiring reference spectra and calibrating the power levels of the pulses.

The efficiency of a pulse or an algorithm is determined by comparing the resultant polarization of the spins with their equilibrium polarization. The polarization is proportional to the integral over the spin's spectral lines.
These spectra were displayed earlier in Figure \ref{fig:TCE_with_spectrum_ch3}. We acquired the reference spectra immediately after the AC experiments to avoid any drift effects. 

The pulse generated by SIMPSON is given in nutation frequency units (Hz), which should be converted to the attenuation of the transmitter in the lab (dB). When the pulse is applied on two channels, both should be calibrated. The carbon channel was calibrated using a 5\ msec selective pulse, generated by SIMPSON, which rotates C2 to the x axis while keeping C1 on the z axis. The pulse was transmitted in different power levels. The power level selected is the one that gave the minimal integral over C1, while the integral over C2 was close to the corresponding integral in the spectrum of a calibrated hard excitation (90$^\circ$) pulse. Knowing the carbon calibration, the power level of the hydrogen in the polarization exchange pulse was calibrated by finding the power which maximizes the polarization transferred to C2. The robustness of the pulses to RF inhomogeneity was observed during the power calibration.
For instance, a deviation of $\pm0.5$~dB (i.e., $\pm5.9$\%) from the optimal carbon and hydrogen power levels, reduced the final polarization of C2 by only 1.6\% (from 3.8 to 3.74).

\section{Results}\label{chap:Results}

\begin{figure}[ht!]
	\centering
	\includegraphics [scale=0.55] {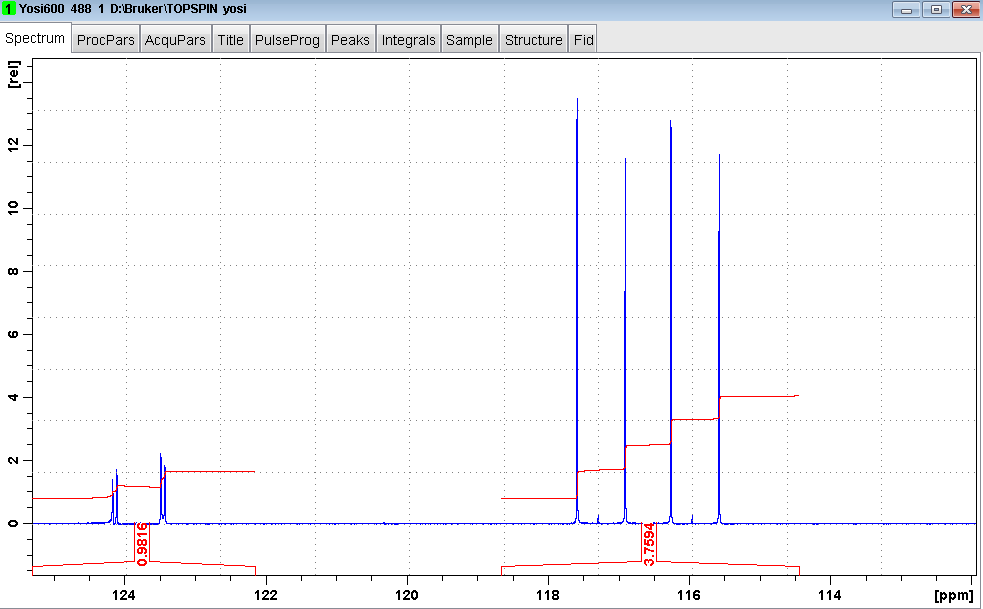}		
			\caption { The spectrum of the carbons after applying polarization exchange, followed by a non-selective  90$^\circ$ pulse. The integrals of C1 and C2 are $0.98\pm0.02$ and $3.76\pm0.02$ respectively, compared to a calibrated \textbf{zg} spectrum.
	}
			
		\label{fig:PolarizationSwap600}
	\end{figure}

 We measured a spectrum corresponding to the PE pulse (followed by a readout pulse) applied onto the equilibrium state. The spectrum clearly exhibits a polarization increase of C2: the measured polarization of C1 and C2 are 0.98$\pm$0.02  and 3.76$\pm$0.02 respectively. The transfer efficiency is therefore 94\%, and the loss of C1 polarization is close to our prediction. The spectrum is displayed in Figure \ref{fig:PolarizationSwap600}. 

\begin{figure} [H]
\centering
\includegraphics [scale=0.7] {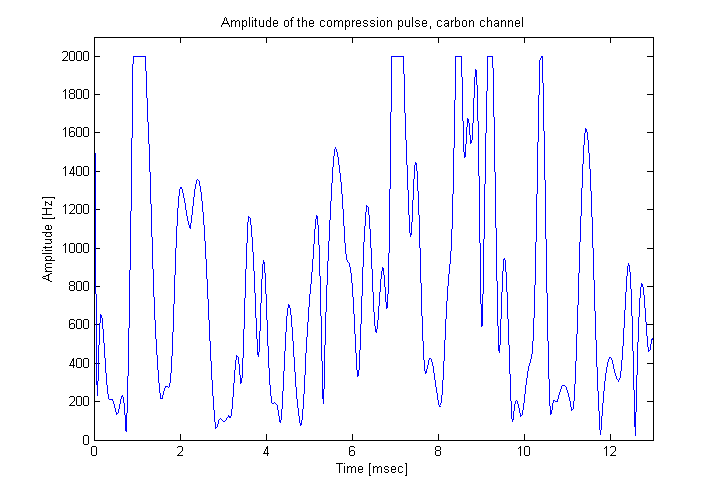} 
\includegraphics [scale=0.7] {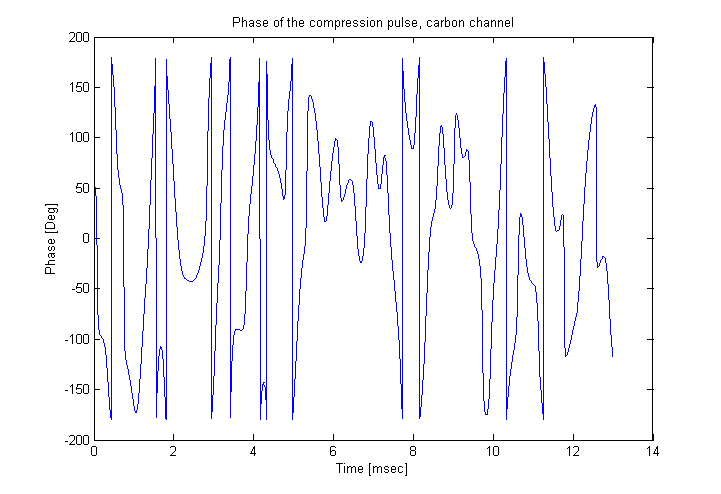} 

\caption [The amplitude and phases of the compression pulse on the carbon channel.] {The shape of the compression pulse on the carbon channel. In the top graph, the vertical axis is the RF amplitude, limited by the maximal power output of the RF transmitter, and the horizontal axis is the time (the total duration of the pulse is 13~msec).  The bottom graph shows phases of the same pulse (in degrees).}
\label{fig:comp_amp_phase}
\end{figure}

\begin{figure} [hf]
\centering
\includegraphics [scale=0.7] {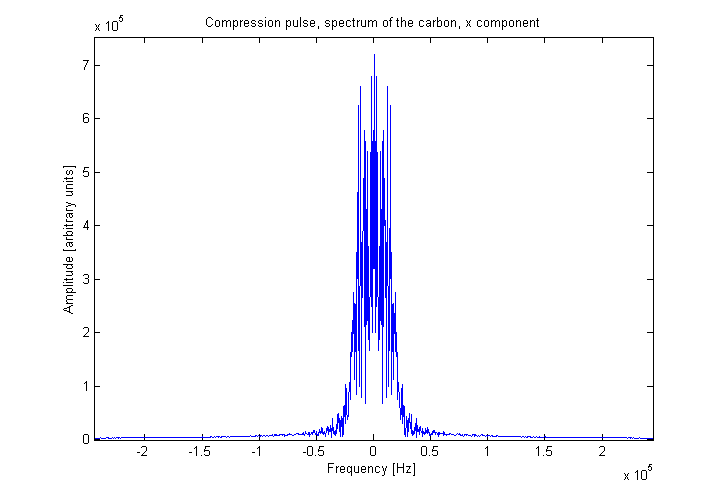} 
\caption [Fourier transform of the x component of the compression pulse applied to the carbons.]{Fourier transform of the x component of the compression pulse applied to the carbons. The spectral width is about 50~kHz.}
\label{fig:comp_amp_shape}
\end{figure}

Figure \ref{fig:comp_amp_phase} shows the phase and amplitude of the part of the compression pulse applied to the carbon channel. Note that the rapid modulation of the amplitude and phase during the pulse are not always feasible by the hardware, thus reducing the pulse efficiency. The Fourier transform of the pulse's x component is shown in figure \ref{fig:comp_amp_shape}. It is difficult to understand from this image how the pulse addresses the two carbons. 

\begin{figure}[!hf]
	\centering
	\includegraphics [scale=0.45] {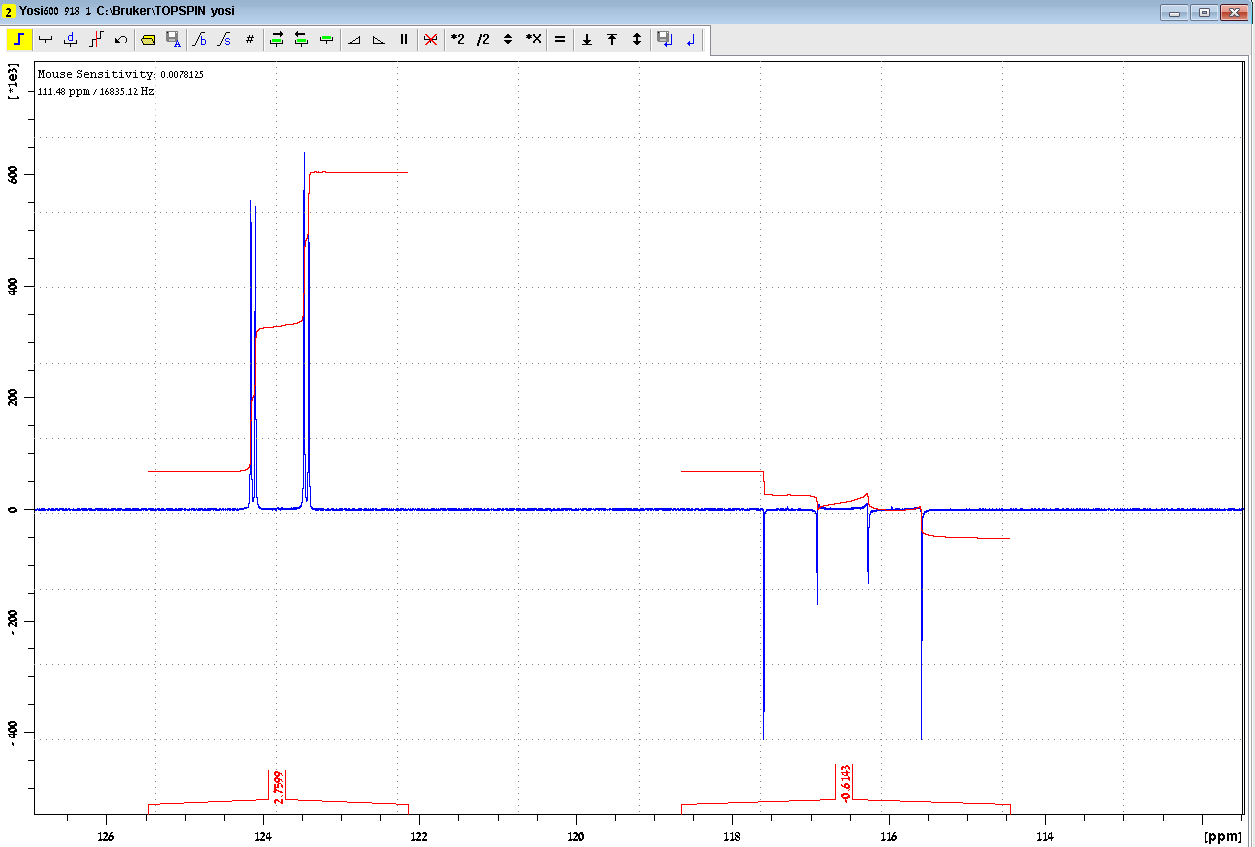}		
			\caption { The spectrum of the carbons after applying a compression pulse, followed by a non-selective  90$^\circ$ pulse. The integrals of C1 and C2 are $2.76\pm0.02$ and $-0.631\pm0.02$ respectively, compared to a calibrated \textbf{zg} spectrum.
	}
			
		\label{fig:CompEq}
	\end{figure}

 Figure \ref{fig:CompEq} shows the spectrum of the carbons following the compression pulse and a readout pulse. The polarizations of C1 and C2 were 2.76$\pm$0.02 and -0.631$\pm$0.02 respectively, while the SIMPSON prediction was 3.00 and -0.73. The efficiency of the compression pulse, which refers to the acquired polarization of C1 was 92\% with respect to the ideal case.

\section{Discussion}\label{chap:Discussion}

In this work  we demonstrated the design and experimental application of two
highly non-trivial quantum gates which are also important building blocks of algorithmic cooling (AC) -- a method for extracting entropy from qubits into the environment.\cite{BMRVV02,FLMR04} The implementation was done by using optimal control in liquid state NMR --- the state-to-state mode of SIMPSON\@. 

GRAPE has been used for NMRQC for the last few years. 
The SIMPSON implementation of GRAPE generated efficient and robust enough pulses. 
The parts missing from SIMPSON - the experimental limitations, as well as dissipation processes and numerical errors, cause a gap between the efficiency of the pulses as expected by the optimization and the observed efficiency. 
The following factors may explain the gap between SIMPSON's predicted pulses efficiency ($>99\%$), and the efficiency observed in the lab which was between 92\% to 94\%.

\begin{enumerate}
\item 
Dephasing - throughout the evolution of the density matrix during the  pulses, the system may experience different dephasing which depends on its instantaneous state. Assuming the dephasing time constant is the spins' average T$_2^*$, 290~ms, the polarization lost during a 6~msec pulse and a 13~msec pulse is roughly 2\% and 4\% respectively.  Although GRAPE can integrate dissipation processes into the state to state optimization, they are not supported by SIMPSON\@. A deeper analysis of the evolution of the density matrix is required for a better estimation.

\item  The probe and the rest of the hardware has limited bandwidth, see for example Refs.~\refcite{BarbaraMartinWurl+91,BLMR+08}, and therefore attenuates and distorts frequencies that are far from the carrier wave. In our case, the distance between the control points of the shaped pulses is roughly a few micro-seconds, which translates to a bandwidth of $\approx$1~MHz.  Therefore, highly modulated pulses may reach lower efficiencies than smooth pulses. 
\item The number of the pulse's control points seems to affect the observed efficiency. The maximal number of point allowed by SIMPSON on our computer are 5000. If we could increase the number of control points then perhaps the efficiency would also increase. 
\end{enumerate}

Possible avenues for further improving pulse fidelity:
\begin{description}
\item \textbf{Pulse design} - An alternative to SIMPSON is introduced in Ref.~\refcite{MSTN10}, an algorithm which generates smooth pulses, easier for the hardware to implement. Another alternative was recently introduced,\cite{2ndOGRAPE+11} an improved version of GRAPE which converges faster by utilizing a quasi-Newton method.  This version (called BFGS-GRAPE) was implemented in Matlab within the Spinach package.  The dephasing during the pulse transmission can be minimized by utilizing algorithms which take dephasing into account, or by giving higher priorities to trajectories of the density matrix with the least dephasing. 
\item \textbf{Hardware} - The fidelity of the pulses can be improved by utilizing a feedback circuit, as described in Ref.~\refcite{RLL09}, which reduces the fidelity degradation due to hardware bandwidth limitation.
\end{description}

Our work paves the way to implementing multiple rounds of algorithmic
cooling in solution,\cite{AEMW14,AtiaMSc} by concatenating PE, compression 
(such as those designed above) and reset steps.
The current work could also be a step towards employing nuclear magnetic resonance quantum computing (NMRQC) and AC in biomedical \iso{C}{13}-magnetic resonance spectroscopy and imaging. AC, and other NMRQC tools, may enhance in-vivo spectroscopy of slow metabolic processes, particularly in the brain, whereby \iso{C}{13}-labeled metabolites (e.g., amino acids) are produced.\cite{EGMW11}

\section{Acknowledgments}
 This work was supported in part by the Wolfson Foundation, and by the Israeli MOD Research and Technology Unit. The work of T.M. was also supported in part by FQRNT through INTRIQ, and by NSERC\@.


\end{document}